# Photodynamic therapy of 4T1 tumors in NOD-SCID mice

**Georgios Kareliotis**[1*], **Stavros Xanthopoulos**[2], **Eleni Drakaki**[1], **Maria Papachristou**[3], **Ioannis Datseris**[3], **Penelope Bouziotis**[2], **Mersini Makropoulou**[1]

[1]Laser Development and Applications Team, Physics Department, School of Applied Mathematical and Physical Sciences, National Technical University of Athens, Zografou Campus, 15780, Athens, Greece

[2]Institute of Nuclear & Radiological Sciences, Technology, Energy & Safety (INRASTES), NCSR "Demokritos", Athens, Greece

[3]Nuclear Medicine - PET/CT Department, General Hospital of Athens "Evaggelismos", Ypsilantou 45-47, 10676, Athens, Greece

*Corresponding author
Georgios Kareliotis
E-mail: gkarel@central.ntua.gr
Laser Development and Applications Team, Physics Department, School of Applied Mathematical and Physical Sciences, National Technical University of Athens, Zografou Campus, 15780, Athens, Greece

**Abstract**
Background: Promising results for mammary carcinoma treatment with photodynamic therapy (PDT) presuppose a careful selection of irradiation light wavelength.
Methods: 4T1 tumors implanted in NOD-SCID mice were treated with Metvix-PDT under 625 nm, 660 nm and their combination light, for a fixed radiant exposure. The therapeutic outcome was assessed through Monte Carlo based computational simulations along with a preliminary *in vivo* study, where fluorescence, size and temperature measurements were conducted.
Results: The light source combination protocol presents great potential, since it results in high cytotoxic products levels and reduced treatment times; while the *in vivo* findings, regarding the harvested tumor mass, also support this hypothesis. The irradiation with 625 nm beam alone presented better results for most of the *in vivo* measured parameters. The mouse treated with only the 660 nm light source had the highest un-photobleached photosensitizer (PS) signal, the lowest body temperature, the heaviest harvested tumor and the lowest estimated concentration of PDT cytotoxic products.
Conclusions: The use of 625 nm irradiation light matches the PS excitation band but is preferable only for treatment of superficial tumors. For deeper laying masses, the simultaneous use of longer wavelengths enhances the therapeutic outcome due to their increased penetration depth.

## 1 Introduction

Photodynamic therapy (PDT) has been used in cancer treatment research since 1978 [1]. It aims to tumor regression and eradication based on three main aspects: light, photosensitizing drug (PS) and molecular oxygen ($^3O_2$). The drug, topically or systemically administrated to the patient, is selectively absorbed by tumor cells. The irradiation that follows with light of appropriate wavelength ($\lambda$) leads to formation of cytotoxic products, which induce cancer cell death via apoptosis, necrosis, autophagy or regulated necrosis (e.g. parthanatos) [2,3]. The combination of the light fluence rate and PS type strongly affect the cell death mechanism [4]. Even though PDT is mostly used in cases of superficial tumors, intravenously injected PSs and fiber optics allow also treatment of deep located cancer [5]. This is advocated by the essential PDT advantage of sparing adjacent healthy tissue, since the PS molecules are mainly concentrated in the malignant area and remain inactivate until illuminated.

Another important advantage of topically-administrated photosensitizing agents is the reduced systemic toxicity as well as the decreased accumulation in organs like liver. On the other hand, the drug has to go through skin layers reducing its final amount, making the exact and fast calculation of its concentration and dosimetry a puzzling procedure. Furthermore, in the effort for accurate dosimetry one should account the rather complex light distribution inside tissues and the even more complicated tumor oxygenation. Light and PS dosimetry in PDT remain open research fields, although plenty of studies are available in literature [6–9].

As it is well known, the light is mainly characterized by its wavelength. Different $\lambda$ values result in different type of interactions with tissues and thus, different light penetration depths. The main tissue parameters that affect penetration are absorption and scattering coefficients ($\mu_a$ and $\mu_s$, respectively), anisotropy factor ($g_f$) and refractive index ($n$). A careful literature overview for the above values reflects the difficulty in gathering all the available data that are necessary to implement an accurate computational PDT simulation.

The sensitizing drug's interactions with the irradiating beam and its pharmacokinetic properties also affect PDT dosimetry. These are quantified in the form of parameters whose precise values are not yet fully known [10], while some of them are discussed in section *"2.3 Simulation model"*. In brief, the appropriate irradiation light efficiently excites the PS molecules from ground to triplet energy state through intersystem crossing procedures. From there they can transfer their energy excess to molecular oxygen and convert it to its singlet, cytotoxic form ($^1O_2$) in a procedure called Type II mechanism [10].

4T1, a highly tumorigenic cell line, produces mammary tumors which can spontaneously metastasize from the primary to multiple distant sites [11]. With breast cancer being the most frequently diagnosed women's malignancy and also a leading cause of cancer death [12], and since 4T1 cell line has been little explored from a PDT perspective, its study deserves the attention of researchers [13,14]. Hence, the main objective of this article is to present a computational simulation on the investigation of the use of different light wavelengths in the PDT treatment of mammary carcinoma in mice. A preliminary *in vivo* study was also implemented in order to assist the *in silico* assessment. The results regarding fluorescence, tumor size and temperature variations are given for all the post-treatment period.

## 2 Materials and Methods

### 2.1 Brief protocol description

PDT for breast carcinoma scenarios (4T1 cells) in NOD severe combined immunodeficient (SCID) mice was investigated. Four different irradiation protocols were incorporated, with the light parameters being presented in **Table 1**. In order to study the effect of each excitation wavelength on the therapeutic outcome, three mice received the same radiant exposure ($H$), which was set at approximately 100 J·cm$^{-2}$ [15,16]. It should be also noticed that fluence rate values ($\varphi$) for 4T1 PDT in literature range up to ~400 mW·cm$^{-2}$ [17] but in this study were limited to ~200 mW·cm$^{-2}$, according to the PS's Summary of Product Characteristics (SmPC) [18] and to each source's, used in the preliminary *in vivo* experiment, available power. The radiant exposure was controlled by adjusting the irradiation time. The control mouse (No. 4) was not irradiated.

### 2.2 Photosensitizer

The PS studied was Metvix cream (Galderma, France), since its lipophilicity offers relatively deep penetration into tumors [19,20]. This second generation PS contains methyl aminolevulinate (MAL), as hydrochloride ($C_6H_{11}NO_3$•HCl) (160 mg/g) and among others, cetostearyl alcohol (40 mg/g), methyl parahydroxybenzoate (2 mg/g), propyl parahydroxybenzoate (1 mg/g) and arachis oil (30 mg/g) [18]. Metvix presents excitation maxima at 405 and 635 nm [10]. The methyl aminolevulinate contained is metabolized to intracellular porphyrins, including Protoporphyrin IX (PpIX), which is highly photosensitive and mainly accumulates in tumor rather in healthy cells [4].

**Table 1** Treatment protocol for each mouse

| Parameter | Mouse No. 1 | Mouse No. 2 | Mouse No. 3[a] | | Mouse No. 4 (Control) |
|---|---|---|---|---|---|
| $\lambda$ **(nm)** | 625 | 660 | 660 | 625 | - |
| $P$ **(mW)**[b] | 47.5 | 157 | 157 | 47.5 | - |
| $\varphi$ **(mW·cm$^{-2}$)**[c] | 60.5 | 199.9 | 199.9 | 60.5 | - |
| $H$ **(J·cm$^{-2}$)** | 99.9 | 102 | 50.3 | 49 | - |

[a] Mouse No. 3 was irradiated simultaneously with the laser and the LED source. Each column represents the values of the relevant source.
[b] At the end of the fiber. The power difference of the two light sources is a result of their maximum P value available.
[c] At the tissue surface.

*2.3 Simulation model*

The simulation was based on an improved version of a former published code [21,22]. The computational therapeutic scheme was assessed via Type II PDT mechanism. Although the mathematical model describing this mechanism is rather complex, it can be simplified in the following equations' set [7]:

$$\frac{d[S_0]}{dt} + \left(\xi\sigma \frac{\varphi([S_0] + \delta)[^3O_2]}{[^3O_2] + \beta}\right)[S_0] = 0, \quad (1)$$

$$\frac{d[^3O_2]}{dt} + \left(\xi \frac{\varphi[S_0]}{[^3O_2] + \beta}\right)[^3O_2] - g\left(1 - \frac{[^3O_2]}{[^3O_2]_0}\right) = 0, \quad (2)$$

$$\frac{d[^1O_2]_{rx}}{dt} - f\left(\xi \frac{\varphi[S_0][^3O_2]}{[^3O_2] + \beta}\right) = 0. \quad (3)$$

In the these equations $[S_0]$ stands for the concentration of ground-state photosensitizer, ($t$) for irradiation time, ($\xi$) corresponds to specific oxygen consumption rate, ($\sigma$) to specific photobleaching ratio, ($\varphi$) represents the light fluence rate, ($\delta$) low concentration correction, $[^3O_2]$ concentration of molecular oxygen, ($\beta$) oxygen quenching threshold, ($g$) oxygen supply rate to tissue, $[^3O_2]_0$ initial concentration of molecular oxygen, $[^1O_2]_{rx}$ concentration of singlet oxygen leading to cell death (apparent reacted singlet oxygen) and ($f$) the fraction of $^1O_2$ interacting with target. In this study the threshold concentration of singlet oxygen leading to cell death $[^1O_2]_{rx,sh}$ was set at 500 μM. The spatial profile of the fluence rate was obtained by a Monte Carlo algorithm (mcxyz.c, version July 22, 2019, created by Jacques S., Li T., and Prahl S., available from: https://omlc.org/software/mc/mcxyz/index.html), for 2 million photons.

The model reproduced a cube of edge length of 2 cm, whose upper 0.3 cm represented air and the rest parietal peritoneum muscle tissue. A 0.5 cm – radius tumor volume was assumed to be at the upper central area of the structure, with the first 0.3cm projecting outside muscle tissue. It also incorporated a healthy skin layer with thickness $d = 220$ μm that included epidermis and dermis, based on literature values [23,24]. Its absorption and scattering coefficients were extracted from Refs. [25,26]. The skin scattering anisotropy factor, which accounts for the scattering angles of photons travelling into tissue, was set at 0.8 and the refractive index at 1.4, based on a former study of ours [22]. The corresponding values for the parietal peritoneum muscle tissue were extracted from Refs. [27,28]. As far as the optical properties of 4T1 cells are concerned, it was not possible to find in literature values corresponding to the irradiation wavelengths used. Therefore, $\mu_a$ and $\mu_s$ tumor values were extracted by interpolation of data available in Refs. [29,30]. They were based on the trendlines of the coefficients' graphs of Ref. [31], and are presented in **Table 2**. The corresponding values for the 405 nm source were extracted from available data of Ref. [26]. Moreover, since *n* is related to protein concentration and because cancerous tissues present higher protein concentration values than healthy ones its value was fixed at 1.54 [22]; while $g_f$ was fixed at 0.9 [22]. Since the PS cream used leads to PpIX formation, its contribution to optical properties of the tumor area was taken into account and its optical properties were extracted from literature [32,33]. For the simulation total absorption and scattering coefficients at tumor site were calculated as: $\mu_a = \mu_{a,4T1} + \mu_{aPpIX}$ and $\mu_s = \mu_{s,4T1} + \mu_{sPpIX}$, respectively [34].

The photochemical parameters used in the computational simulations correspond to those of 5-aminolevulinic acid PpIX (ALA-PpIX), as MAL is demethylated by esterases to ALA [35]. They were extracted from Refs. [7,10,36] and are presented in **Table 2**. As the average administered PS mass to each mouse was 30 mg (corresponding to $m = 4.8$ mg of methyl aminolevulinate, as hydrochloride), it formed a cylinder with height $h \approx 0.1$ cm and radius $r \approx 0.8$ cm, and because the molar mass ($M_r$) of methyl aminolevulinate hydrochloride is 181.62 g/mol [37], the initial PS concentration on skin surface $[S_0]$ was estimated at 131 mM, using **Equation 4**:

$$[S_0] = \frac{m}{M_r \pi r^2 h} \quad (4)$$

A correction factor of 0.1 was applied to the $[S_0]$ value to take into account the worst case scenario of Parafilm removal by the mice during the 3 h dark interval for PS absorption, as well as resultant PS loses (e.g. PS remains on the glove of application and on mice hair). Moreover, according to the PS's SmPC [18], *in vitro* studies in human skin revealed a dermal depot containing 4.9% of the radiolabelled MAL administered dose, although, as stated "in humans, a higher degree of accumulation of porphyrins in lesions compared to normal skin has been demonstrated with Metvix cream". Therefore, in this study, after the dark period and the area cleaning with saline, the PS absorption through skin was set at 5% of the surface concentration. Furthermore, to account for the PpIX build up during the dark period as a function of depth, the trendline of the relevant data available in a 2016 study of Campbell et al. [32] was extracted and used. Finally, the PpIX molecules were assumed to have accumulated only in tumor tissue.

All the above mentioned calculations were implemented through scripts coded in MATLAB software (The MathWorks, Inc., Natick, MA, United States), using a desktop computer with an Intel® Core™ i7 – 4790 @ 3.60 GHz processor and 16 GB of RAM.

**Table 2** Optical and photochemical parameters of the simulation model.

| Layer | Parameter | Optical parameters [a] | | | Photochemical parameters [b] | |
| --- | --- | --- | --- | --- | --- | --- |
| | | Value @ 405 nm | Value @ 625 nm | Value @ 660 nm | Parameter | Value |
| Skin | $\mu_a$ | 2.86 cm$^{-1}$ | 0.33 cm$^{-1}$ | 0.20 cm$^{-1}$ | $\varepsilon$ | 0.003 (cm$^{-1}$ μM$^{-1}$) |
| | $\mu_s$ | 113.9 cm$^{-1}$ | 166.4 cm$^{-1}$ | 165 cm$^{-1}$ | $\xi$ | 3.7 · 10$^{-3}$ (cm$^2$ mW$^{-1}$ s$^{-1}$) |
| | $g_f$ | | 0.8 | | $\sigma$ | 9 · 10$^{-5}$ (μM$^{-1}$) |
| | $n$ | | 1.4 | | $\beta$ | 11.9 (μM) |
| | $d$ | | 220 μm | | $f$ | 1 |
| | | | | | $\delta$ | 33 (μM) |
| 4T1 tumor | $\mu_a$ | 3.89 cm$^{-1}$ | 0.94 cm$^{-1}$ | 0.76 cm$^{-1}$ | $g_0$ | 0.7 (μM s$^{-1}$) [c] |
| | $\mu_s$ | 68.5 cm$^{-1}$ | 86.2 cm$^{-1}$ | 98.0 cm$^{-1}$ | $[^3O_2]_0$ | 60 (μM) |
| | $g_f$ | | 0.9 | | $[^1O_2]_{rx,0}$ [d] | 0 |
| | $n$ | | 1.54 | | $[S_0]_{corrected}$ [e] | 655 (μM) |
| | $d$ | | 1 cm | | $[^1O_2]_{rx,sh}$ | 500 (μM) |
| PpIX | $\mu_a$ | 1.88 cm$^{-1}$ | 0.06 cm$^{-1}$ | 0.0 cm$^{-1}$ | | |
| | $\mu_s$ | 115.0 cm$^{-1}$ | 66.7 cm$^{-1}$ | 66.7 cm$^{-1}$ | | |
| | $g_f$ | | 0.9 | | | |
| | $n$ | | 1.38 | | | |
| Parietal peritoneum muscle tissue | $\mu_a$ | 3.0 cm$^{-1}$ | 2.2 cm$^{-1}$ | 2.2 cm$^{-1}$ | | |
| | $\mu_s$ | 36.4 cm$^{-1}$ | 20.5 cm$^{-1}$ | 19.6 cm$^{-1}$ | | |
| | $g_f$ | | 0.8 | | | |
| | $n$ | | 1.4 | | | |

[a] $g_f$, $n$ and $d$ values apply for all wavelenght values of each layer.
[b] Obtained from Ref. [10,36].
[c] Obtained from Ref. [7].
[d] Initial concentration of singlet oxygen leading to cell death.
[e] PS concentration finally absorbed (after correction factors).

*2.4 Animals and tumors*

For the preliminary *in vivo* part of the study twelve-week-old female NOD-SCID mice (n = 4) were acquired from the Experimental Animal Colony of the Institute of Biosciences & Applications of National Center for Scientific Research "Demokritos". They were housed two per cage and were maintained and treated according to the EU guidelines and ethics. The average mouse weight at the PDT day was 22.37 g, measured with a digital balance (440-33N, KERN & SOHN GmbH, Balingen, Germany). The animals were sacrificed at day 12, when the control mouse spontaneously died, by putting them in an airtight chamber filled with Iso-Vet (Isoflurane 1000 mg/g Inhalation vapor).

Breast carcinoma 4T1 cells were obtained by National Center for Scientific Research "Demokritos". They were prepared in a 0.25 mL suspension containing 1·10$^6$ cells that was inoculated subcutaneously in the left flank of each mouse. The tumors were allowed to grow until their diameters reached approximately 1 cm.

*2.5 Fluorescence and temperature imaging*

The fluorescence spectrum of the treated area was acquired (i) before the PS application, (ii) just before the PDT and (iii) immediately after the treatment procedure. A 405 nm (8 mW) LED source was used as excitation source (M405FP1, Thorlabs, Germany), since the maximum of the PpIX's Soret band is located at 405 nm. The light was collected by a 600 μm core optical fiber (QP600-2-SR-BX, Ocean Optics), attached to a spectrometer (USB 4000, Ocean Optics). A 600 nm longpass filter (FEL0600, Thorlabs, Germany) was used to eliminate the 405 nm light along with unwanted spectrum areas. The collected spectrum was processed using SpectraSuite software (Ocean Optics), with integration time set at 100 ms and the background signal was subtracted.

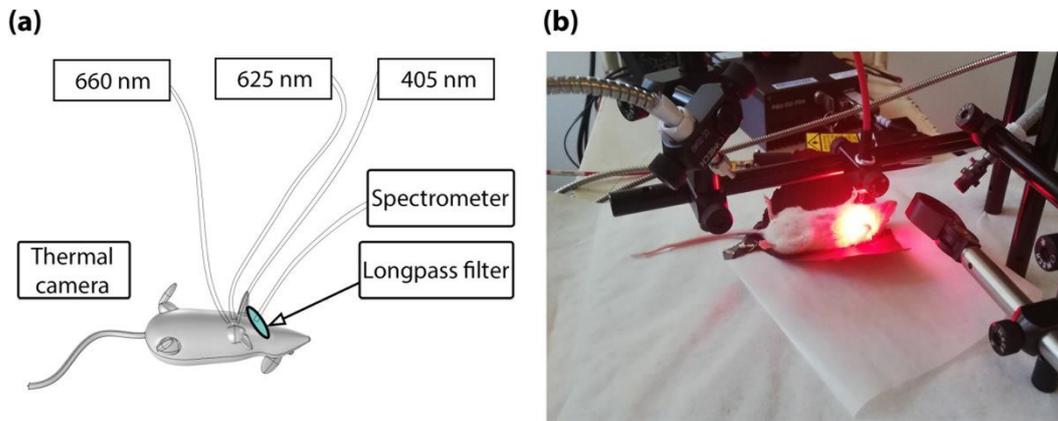

**Figure 1.** The experimental setup used: **(a)** Schematic representation of the main parts. **(b)** The optical fiber with the metallic housing (on the left of the image) guided the 660 nm beam. The one in the middle (above mouse) guided the 625 nm or the 405 nm beam (depending on protocol). At the right of the image the collection optical fiber and the longpass filter, which were placed as close as possible to the mouse during fluorescence measurements, can be seen.

Surface temperature measurements of the mice's abdomen and their left flank were conducted using a thermal camera (ToughCam EL S, Infrared Cameras Inc.). The setup is presented in **Figure 1**.

*2.6 Treatment procedure*

The hair at the tumor area was carefully shaved the day before treatment. Just before applying the PS, each mouse was weighed and the fluorescence spectrum of the target area was acquired. A thin layer (approximately 1 mm thick) of PS, corresponding to ~30 mg was applied topically at the left flank of each mouse, in excess of the visible tumor area, in order to cover a 1.5 mm margin. The area was covered with Parafilm (PM 996) which was attached by surgical tape (Micropore™, 3M™) and then the mouse was kept for 3h in its cage. Cages were covered with black fabric to prevent unwanted photosensitization. Afterwards, the mouse was intraperitoneally injected with a solution of ketamine/xylazine (hydrochloric ketamine 100 mg/mL, 10% and hydrochloric xylazine 23.32 mg/mL, 5%), at 10 mg/kg of body weight.

The cancerous area was irradiated using a 625 nm LED (M625F2, Thorlabs, Germany) (mouse No. 1) and a 660 nm diode laser (PSU-III, CNI Optoelectronics Tech. Co., China) (mouse No. 2). One mouse (mouse No. 3) was subjected to combined therapy, with both light sources illuminating it simultaneously, in a 50-50 light dose separation. The light sources' power values (P) were obtained using a power meter (PM100A, Thorlabs, Germany) with a Si Photodiode detector (S121C, Thorlabs, Germany).

*2.7 Data analysis*

Treatment evaluation was performed by examination of the growth curves of the PDT treated and the control mice. The measurements were conducted by setting the jaws of a caliper (with 0.01 mm accuracy) around the largest and smallest tumor diameters observed. The tumor area and the tumor volume were estimated using **Equations 5** and **6**, respectively.

$$Area = \pi \cdot \frac{Diameter_{max} \cdot Diameter_{min}}{4} , \quad (5)$$

$$Volume = \frac{Diameter_{max} \cdot Diameter_{min}^2}{2} . \quad (6)$$

At day twelve after treatment the mice were sacrificed, the tumors were extracted and their masses were recorded.

**3 Results**

The computational simulation revealed the differences arising from the penetration depth of each light source. As it can be seen from **Figures 2(b-d)**, longer wavelengths offer higher fluence rates in deeper planes; and as a result deeper lying tissue is able to absorb significant higher amounts of energy (see **Fig. 2(b) and (f)**). The skin and tumor $\mu_s$ values were accounted to be responsible for the increased $\varphi$ values recorded on the upper layers of the model, where transmmited and scattered photons coexist. In the 660 nm light case $\varphi$ exceeded superficially the thershold of 200 mW·cm$^{-2}$ in order to achieve high enough values inside the tumor volume.

The variations of cytotoxic produced molecules at the end of PDT irradiation are presented in **Figure 3(a)**. The highest $[^1O_2]_{rx}$ was recorded for the 625 nm light protocol and the lowest for the 660 nm one. Although the total radiant exposure was in all cases the same (approximately 100 J·cm$^{-2}$), the $[^1O_2]_{rx}$ fluctuated by more than a threefold.

The combined light sources protocol applied to mouse No. 3 resulted in intermediate $[^1O_2]_{rx}$ values. One case scenario was implemented in **Figure 3(b)**, where two equal irradiation time values were applied for the two different light sources. Despite the lower penetration depth of the 625 nm beam, its higher, compared to the other beam's, absorption coefficient resulted in ~50% increased $(^1O_2)_{rx}$ production at the first 0.5 cm of the tumor (for irradiation time of 820 s). On the contrary, beyond that point, the higher penetration depth and fluence rate of the 660 nm beam begun to compensate for its lower $\mu_a$ value. Moreover, the increased irradiation time (1650 s) with the 660 nm light produced the same amount of $(^1O_2)_{rx}$, denoting a possible treatment saturation, as far as the specific parameters' values are concerned. Another case scenario, presented in **Figure 3(c)**, studied three possible combinations of light sources, in order to achieve total radiant exposure of ~100 J·cm$^{-2}$. The 75% - 25% split of 625 nm – 660 nm beams resulted in 3% higher $[^1O_2]_{rx}$ values than those of irradiation with 625 nm only, while treatment time was reduced in half. The corresponding increase for the reverse protocol (25% - 75%) and the sole use of 660 nm was 19%. The graph was deliberately not incorporated into **Figure 3(a)** for clarity reasons. Finally, **Figure 3(d)** presents the irradiation time needed for the protocols of **Figures 3(a, c).**

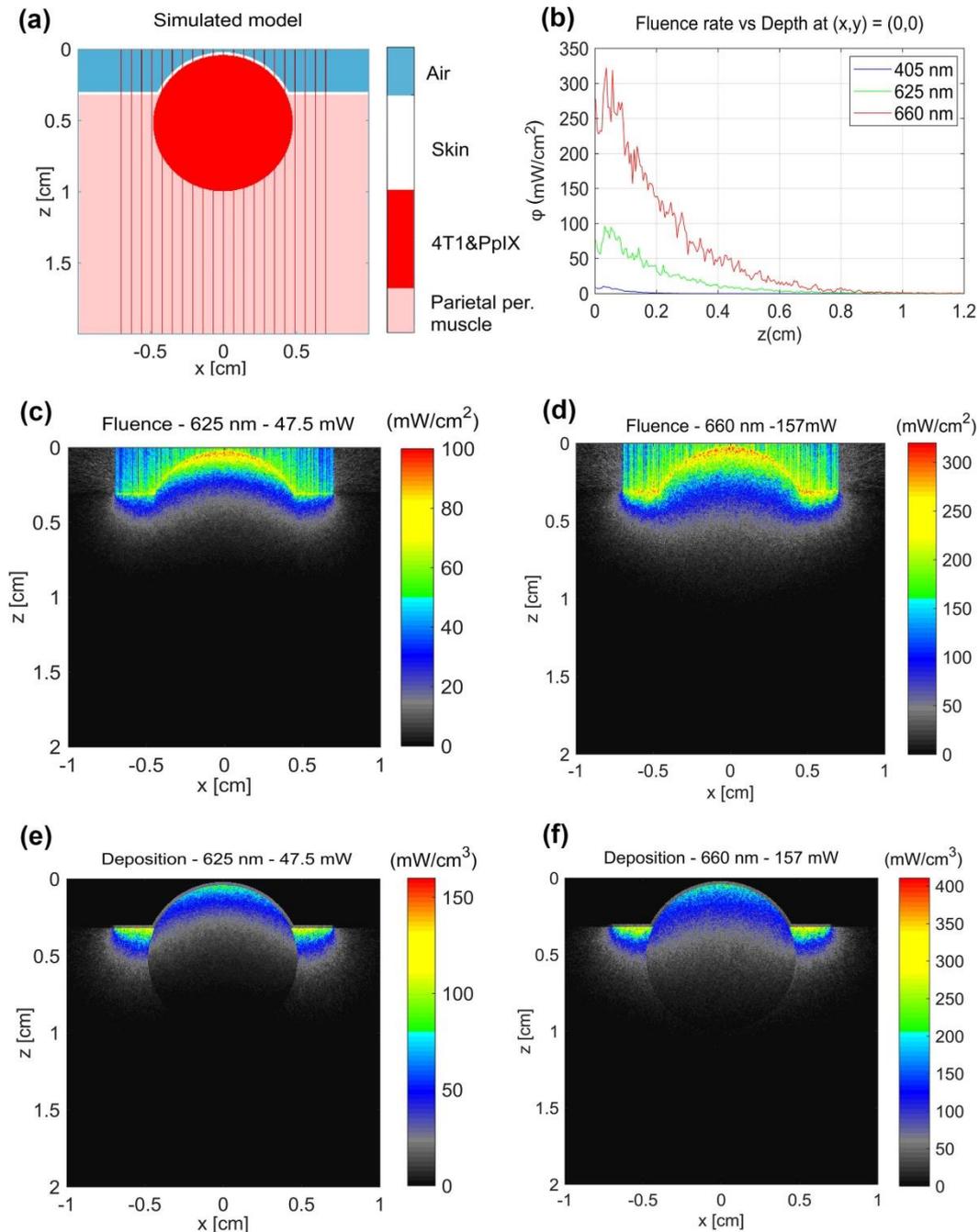

**Figure 2. (a)** Planar view of the simulated model, where the vertical red lines represent the beam's path. **(b)** Spatial profile of fluence rate versus depth for the fluorescence and irradiation beams, for the first 1.2 cm. **(c, d)** Spatial profiles of beam fluence rate and **(e, f)** energy deposition versus depth for the irradiation beams.

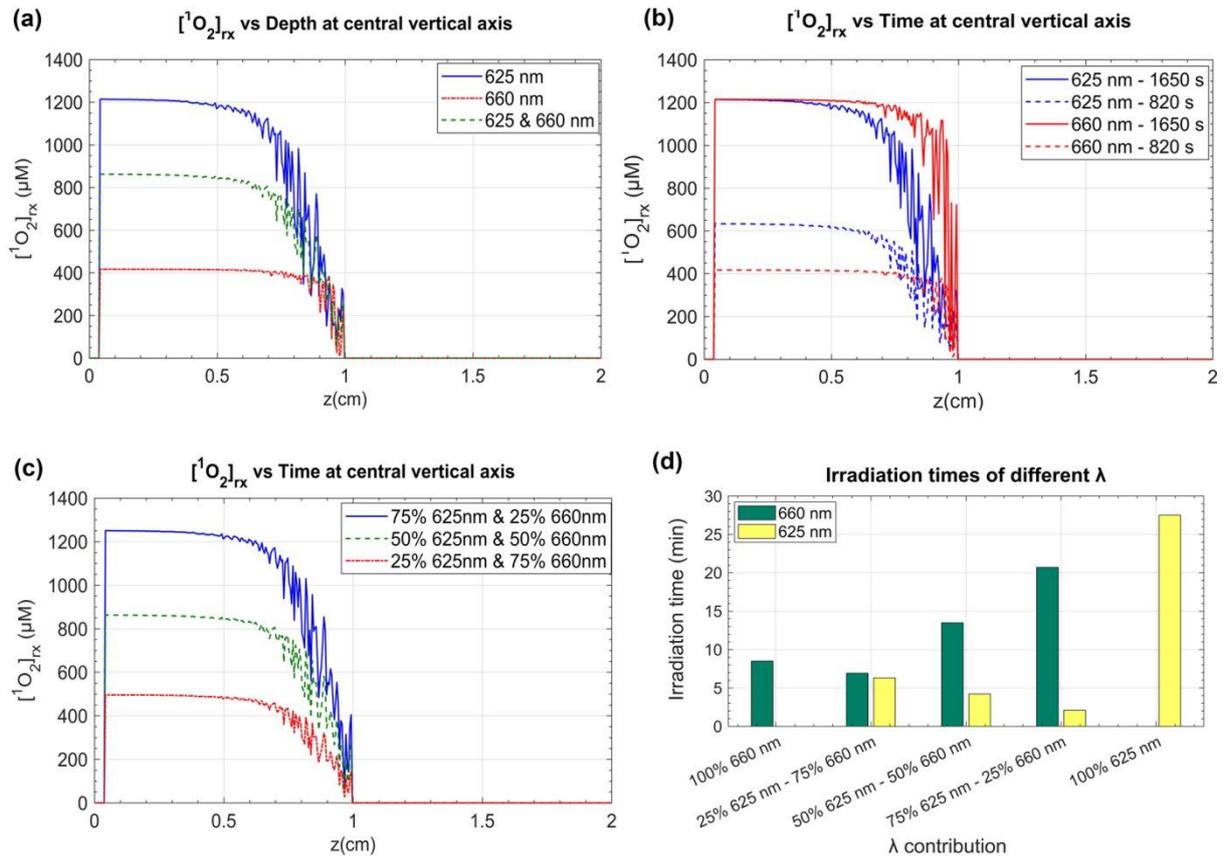

**Figure 3.** (a) Comparison of the $[^1O_2]_{rx}$ produced by irradiation with 625 nm, 600 nm and their combination, for $H \approx 100$ J·cm$^{-2}$. (b) Effect of irradiation time on $[^1O_2]_{rx}$ produced for 625 nm and 660 nm. (c) The effect on $[^1O_2]_{rx}$ of three different combinations of the light sources, in terms of percentage contribution on total radiant exposure of ~100 J·cm$^{-2}$. The 50% - 50% case corresponds to the irradiation conditions of mouse No. 3. (d) Irradiation time needed for protocols of (a) and (c).

The fluorescence spectrum of each mouse was collected after the 3h period of PS uptake, the tumor area wash with sodium chloride and just before irradiation. As shown in **Figure 4(a)** the mouse that was to be irradiated with the 660 nm source presented the lowest intensity signal, while the peak fluorescence of the other two mice (No. 1 and No. 3) was at similar levels. The main three peaks were observed at 604, 636 and 704 nm with minimum variations. **Figure 4(b)** presents the acquired fluorescence spectra of the mice No. 2 and No. 3 right after PDT. The corresponding spectrum of mouse No. 1 is not shown as the data were corrupted. **Figure 4(c)** is indicative of the differences in the fluorescence spectra of an untreated area, an area with accumulated PpIX molecules and the same area after PDT (with dotted blue, continuous red and dashed green lines, respectively; as shown in the online colored version). The pre-PS administration curve presented topical maxima at 604, 622, 750 and 774 nm. After the PS administration the topical maxima were located at 604, 636 and 704 nm, for both pre- and post-irradiation. The main difference between the two latter was the fluorescence intensity, which was more than a twofold for the pre-irradiation case, regarding the main maximum wavelength observed (636 nm). Area and volume measurements of the tumor region are presented in **Figures 5(a)** and **(b)**. The percentage tumor growth is higher after PDT with the combination of 625 and 660 nm light than that of 625 nm and the 660 nm light alone. Interestingly, the tumor of the control mouse showed the lowest increase rate.

Temperature measurements of the mice's tumors and the abdomens were conducted throughout the experiment. As shown in **Figure 5(d),** the tumor area presented higher values than healthy tissue, although as days were passing by this difference tended to minimize. The temperature of the control mouse and the one treated with the combination of light sources (mice No. 4 and No. 3, respectively) was the highest observed (37.3 and 39.3°C, respectively) at day 7. As far as the mouse No. 2 is concerned, it constantly presented lower temperature values than the others.

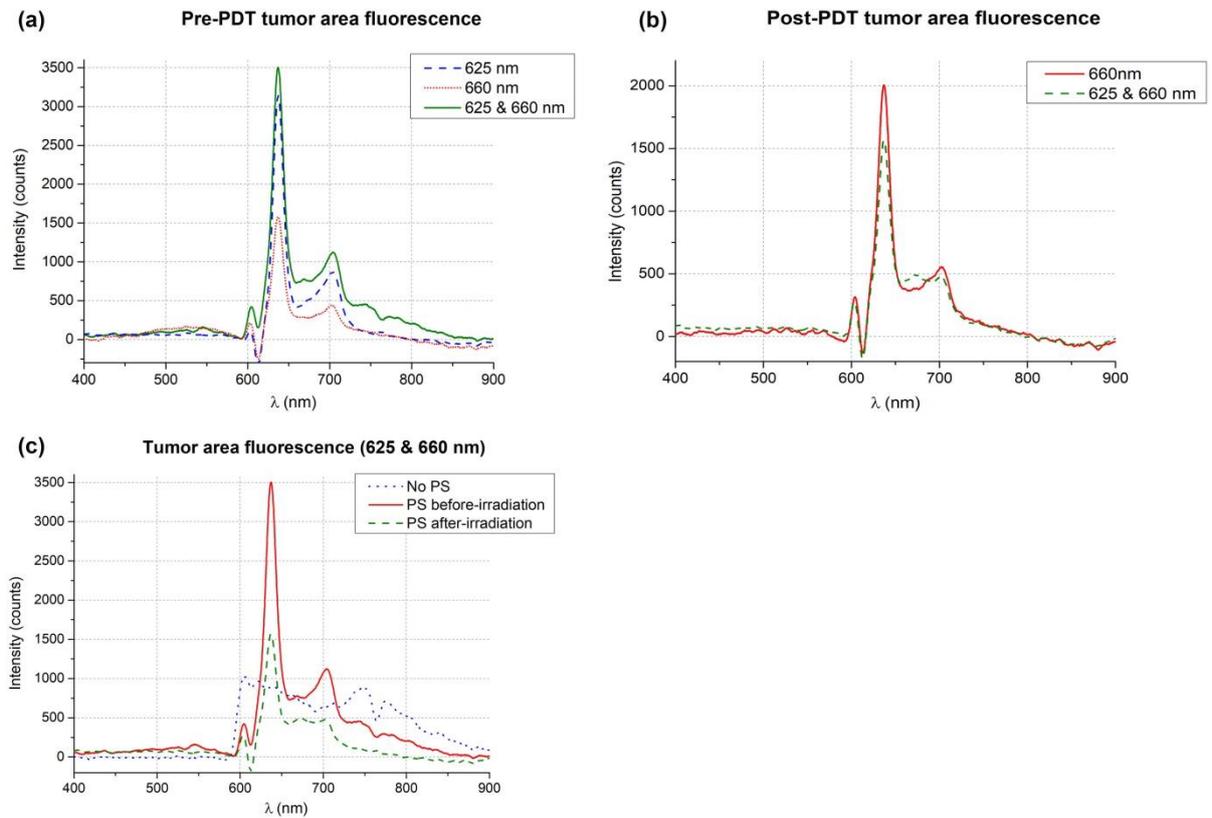

**Figure 4.** Fluorescence spectra of the areas to be treated just before **(a)** and right after **(b)** irradiation with each light source (the spectrum corresponding to the 625 nm right after irradiation was corrupted and is not shown). **(c)** Fluorescence spectra of the treated area: before photosensitizer application (dotted line), just before (continuous line) and right after PDT treatment (dashed line) with both 625 and 660 nm light sources.

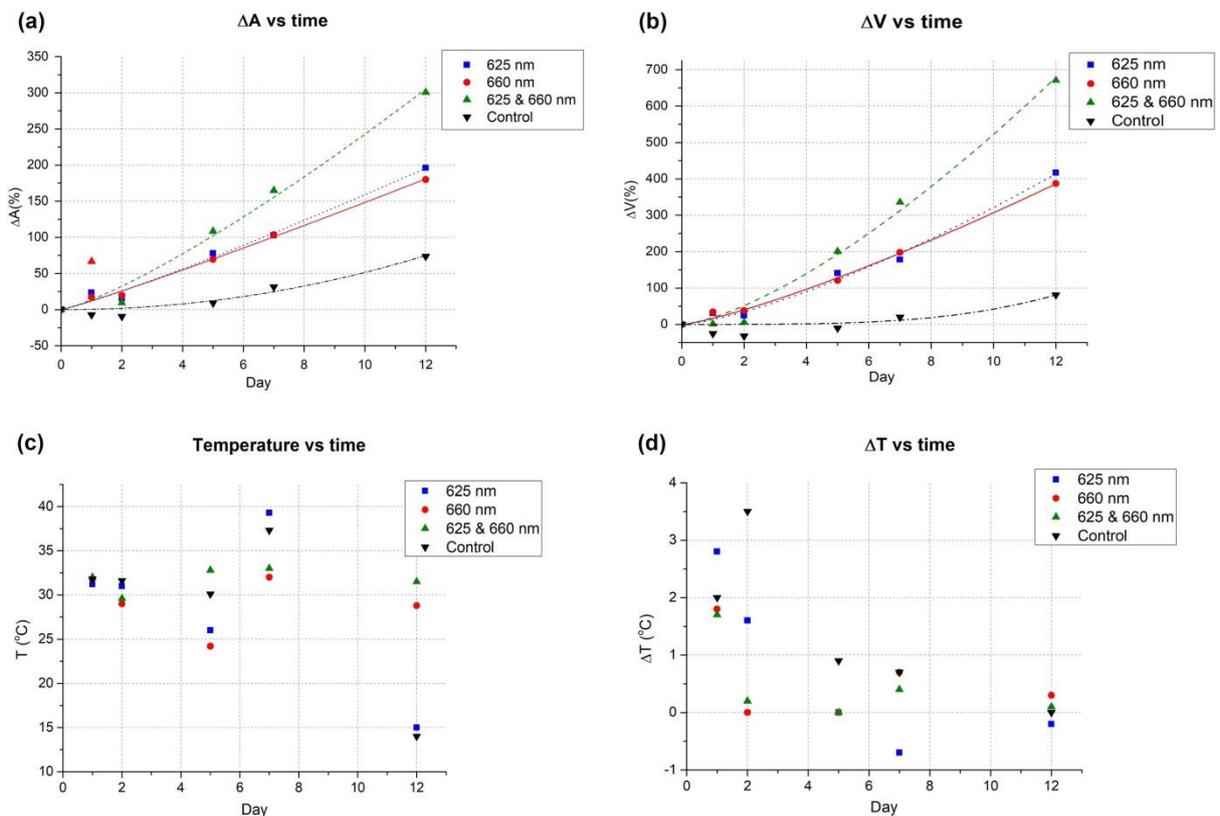

**Figure 5.** Tumor growth expressed as change (%) of **(a)** area at the middle of the tumor and **(b)** volume, for each PDT scheme. **(c)** Tumor temperature throughout protocol and **(d)** temperature variation between tumor and abdomen.

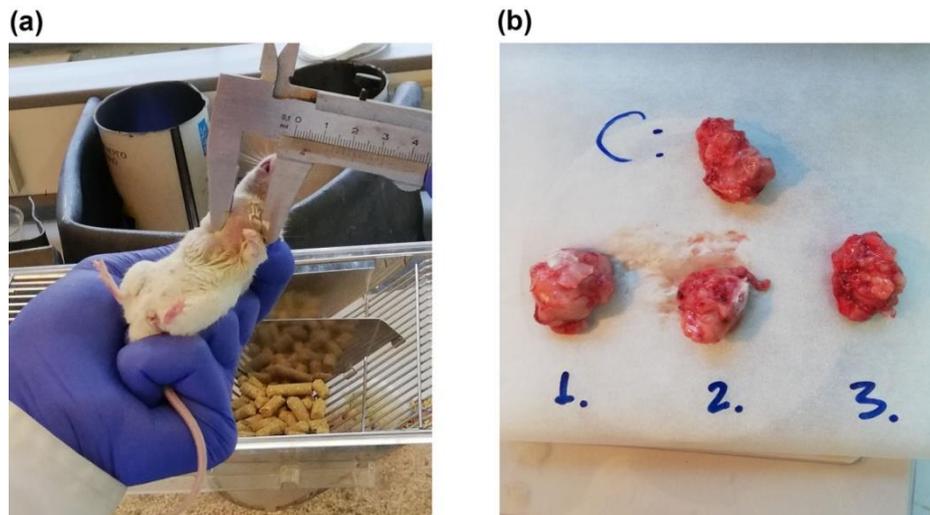

**Figure 6.** **(a)** Tumor size measurement and **(b)** the extracted tumors of the 3 PDT treated mice (numbered sequentially) and of the control mouse (denoted as "C").

Day 12 was considered to be the end of the experiment since the control mouse spontaneously died; and therefore the rest of the animals were euthanized, and the tumors were extracted and weighed. At that day, mouse No. 1 had to be subjected to resuscitational procedures, with its temperature being significantly lower than that of the two alive mice (see **Fig. 5(c)**). At that timepoint the tumors had begun to infiltrate the leg area and also to cover the humerus, while they were consistent in texture and color (see **Fig. 6**). The tumor mass values presented small variations, with the lighter tumor being the one treated with the combination light sources protocol (4.26 ± 0.01 g). Nevertheless, the same value applied also for the tumor of the control mouse. The heavier tumor was that of mouse No. 2 (5.32 ± 0.01 g), while the one of mouse No. 1 weighed 4.69 ± 0.01 g.

**4 Discussion**

In this study Metvix was selected as the photosensitizing agent. It is an affordable, easy to use PS media that does not need to be intravenously injected and therefore presents fewer side effects for the animals. Although the drug's therapeutic indications are focused on the treatment of cancer cases such as actinic keratosis (AK), superficial and nodular basal cell carcinoma (BCC) and squamous cell carcinoma (Bowen´s disease) [18], published studies have also examined its use in mammary adenocarcinoma cells (MDA–MB–231 cells) [16] but to the best of our knowledge, none for the treatment of 4T1 tumors.

The importance of appropriate wavelength selection and irradiation time in $(^1O_2)_{rx}$ production and hence, in treatment outcome is denoted in **Figure 2**. Both 625 nm and 660 nm are within the range of excitation wavelengths proposed by the PS manifacturer [18] but with different efficacy and $\varphi$ values. Although the 625 nm light has lower penetration depth than 660 nm, the corresponding $[^1O_2]_{rx}$ was significantly higher, especially in the first tumor's millimeters. This was attributed to two main factors; even though the total radiant exposure was the same in all mice. Firstly, to the higher value of PpIX's $\mu_a$ at 625 nm than at 660 nm, and secondly, to the more than a threefold longer irradiation time between the mouse treated with 625 nm and the one treated with 660 nm, allowing the tumor's periphery reoxygenation (the reader should bear in mind that its core is hypoxic [38]). The lowest cytotoxic oxygen production was observed for the 660 nm irradiation, where the corresponding $[^1O_2]_{rx}$ was lower than 500 μM. The mouse treated with that protocol presented also the lowest tumor temperature (see **Fig. 5(c)**) and the heaviest extracted tumor. As it can be deducted from the results of the computational simulations presented in **Figure 3**, the lower absorption of 660nm could be compensated by increased irradiation time, although this presupposes prolonged anesthesia time. In clinical practice this could be unconvinient for the patient (PDT induces skin burning sensation) but also time and money consuming for the clinic. At this point, it has to be noted that increased treatment times do not necessarily lead to increased $[^1O_2]_{rx}$ values, as the amount of available PS is not infinite (see **Fig. 3(b)**) and hence, is going to affect PDT efficacy. Additionally, the accumulating photobleached PS molecules act as a barrier for penetrating light, reducing further $(^1O_2)_{rx}$ production. Therefore, an effective combination of 625 nm and 660 nm light should be able to enhance the therapeutic outcome (see **Fig. 3(c)**). The PDT part with the 625 nm light takes advantage of the higher $\mu_a$ value, while the 660 nm part allows treatment of deeper tumor layers (see **Fig. 3(d)**). The combination light sources protocol was further supported by the *in vivo* results, since the corresponding mouse had the lowest extracted tumor's mass, even though it presented the highest rate of tumor area and volume increase (see **Fig. 5(a) and (b)**). Moreover, this mouse presented low tumor and body temperature variations throughout experiment.

Fluorescence spectroscopy plays significant role in non-invasive cancer diagnosis but also in quantification of photosensitizer's concentration. As it is well known, the main cytotoxic product in MAL-PDT is PpIX and hence, its detection is widely used in relevant studies [39,40]. The fluorescence spectra collected just before PDT treatment (see **Fig. 4(a)**) are consistent with the PpIX fluorescence spectrum found in literature [41,42]. They include the main peaks of ~635 and ~705 nm, making a rough correlation between fluorescence intensity and PS concentration possible. The almost half maximum intensity value recorded from mouse No. 2 should not be considered as an absolute value but rather as a quality result; since the tip of the collection fiber was further from the skin than at the rest of the setups, due to technical reasons. On the other hand, that mouse presented the heaviest extracted tumor, denoting a reasonable and expected correlation between the PpIX concentration and therapeutic outcome. The latter is also supported by the increased post-PDT PpIX fluorescence of mouse No. 2, showing its decreased photobleaching and hence, the lower $(^1O_2)_{rx}$ production. Unfortunately, the relevant post-PDT data of mouse No. 3 were not available, as the setup was accidentally violently moved. The PDT outcome is highly correlated to the presence of PpIX, whose concentration fluctuates during treatment. The reader should have in mind that it is produced by MAL conversion but at the same time it is photobleached and biologically removed and hence, its absorption coefficient changes during PDT [43]. Moreover, the cell density also affects treatment's efficiency, through the cellular PpIX distribution [44].

The *in vivo* study was in agreement with the computational model, as the treatment delayed the tumor growth in a manner in agreement with literature [16]. A comparison of the preliminary *in vivo* results with that of Sutoris *et al.* [16] reveals the same behavior of the tumor volume, i.e. a slight increase (day 1) followed by a volume decrease and tumor regrowth. The different time scale (faster in our case) is possibly attributed to the higher aggressiveness of the 4T1 cells used compared to the mammary adenocarcinoma cells of the Sutoris *et al.* study; and to the limited action of superficially applied photosensitizing agents. The control mouse received PS but no light and hence, no effect on tumor growth was expected [16]. It should be noted that the 405 nm light used for fluorescence measurements was not able to induce any PDT anti-tumor effects, due to its low penetration depth (see **Fig. 2(b)**) and its low fluence rate. The fact that the area, volume and mass measurements showed no significant tumor size differences between the control and the treated mice cannot be evaluated in a statistically significant manner, due to the small mice sample. Nevertheless, studies with similar results do exist in literature [26,29], basing their observations mainly on the high blood vessel compression and the hypoxic conditions within the 4T1 tumor. In addition and according to the *in silico* part of the study, the main necrotic area was expected to be on the upper part of tumor volume, considering the superficial irradiation setup. Therefore, the minimum treatment impact to the tumor's periphery (that was mechanically measured) did not come as a surprise. On the contrary, the fact that the untreated mouse died, while most of the rest did not present deteriorated clinical status can account as a successful treatment outcome.

The mechanically obtained measurements of the maximum and the minimum tumor diameters were based on **Equation 6**, which is widely used in literature to estimate tumor volume [16,45]. In this experiment though, the PS and the light delivery were applied from the skin overlaying on the tumor, making the treatment more superficial than whole-tumor based, and thus not uniform, and the use of **Equation 6** not ideal. In order to improve our estimations, the middle horizontal plane area was calculated using **Equation 5**. As the results showed, the latter way seems to be more precise for this type of therapy, since it presented lower divergence than the former as far as the tumor mass is also concerned. Nevertheless, the extracted tumor mass measurements revealed that even more accurate and sophisticated imaging methods are needed for this kind of study (e.g. magnetic resonance imaging – MRI), as the best treatment outcome (the one of mouse No. 3) presented the highest tumor growth rate, and the worst (the one of mouse No. 4 that died first) the lowest tumor growth rate.

The *in vivo* protocol of this study also included regular temperature measurements of both the tumor and the body (abdomen) of the mice before and after PDT. Unfortunately, the literature is poor in relevant post-PDT data since most published studies focus on the photothermal effects during and not after therapy and hence, a comparison with our results was not possible [46,47]. Nevertheless, temperature monitoring is a high value procedure as it can reflect the mouse's condition in a rapid and non-invasive manner [48,49]. Furthermore, temperature heterogeneities are expected between the tumor and adjacent tissue, since tumors present abnormal perfusion and metabolic rates [50]. This fact was also observed in the present study where the tumor temperature presented higher values than its surrounding area, in accordance with other 4T1 studies in literature [51,52], probably attributed to local inflammatory response [53]. At sacrifice day, the temperature of mouse No. 1 was approximately 50% lower than the one of mice No. 2 and No. 3, attributed to decreased metabolic rhythms that possibly denoted a decline in its vital functions [54]. The decreasing temperature difference between healthy and tumorous tissue following treatment was attributed to the vascular shutdown, as the limited blood perfusion into the tumor lead to lower thermal loads.

Conventional ways to increase the PDT therapeutic outcome include the selection of PSs with high absorption at the wavelength used, as well as increased $[S_0]$ and irradiation time. However, toxicity, pain, time and cost issues are raised. Moreover, most approved and clinically available PSs present absorption maxima below the infrared (IR) part of the electromagnetic spectrum, where penetration depth is limited. Consequently, the need of alternative methods to enhance PDT efficacy leads to application of novel modalities. Nanoparticles (NPs) can play a crucial role in drug delivery and selectivity [55]. A specific NP type, namely upconversion NPs, has the ability to emit light of shorter wavelength than the one used for its irradiation [56]. Hence, deep penetrating beams (i.e. IR light) can be used to indirectly excite photosensitizing agents that have much lower excitation bands. The combination of treatment modalities also shows upscaled therapeutic outcome. The synergistic effects of radiotherapy and proton therapy schemes along with PDT offer promising results and an emerging research field [57,58]. Other, less sophisticated ways to enhance tumor response do exist, as for instance administration of low vitamin D doses [59]. Additionally, housing temperature has also been correlated to treatment outcome, as cold-stress can activate physiological changes that affect tumor growth [60].

## 5 Conclusions

This *in silico* study, accompanied by some preliminary *in vivo* results, presented the impact that different light protocols have on PDT of 4T1 tumors. The optical properties of the photosensitizing agent along with the irradiation parameters strongly affect the therapeutic outcome. Moreover, the conventional equations for tumor volume estimation seem to be insufficient, when PSs are superficially applied. The simultaneous acquisition of fluorescence data, in addition to temperature measurements are seen to assist the treatment prognosis. Computational simulations, although with some assumptions, are supported by the fluorescence and temperature data that show decreased treatment effect when using light with a mismatch to the maximum excitation wavelength of the PS, even if high fluence rates are used. The choice of appropriate excitation light that matches the absorption coefficient of the PS is crucial but presents limitations if it has low penetration depth. In this study's case, the best results arose from the appropriate combination of light sources (75% - 25% split of 625 nm – 660 nm beams) that seems to be more sufficient than solely manipulating treatment parameters, such as irradiation time. As a result, we strongly believe that this protocol deserves further investigation, as it has the potential to attribute to the enhancement of therapeutic outcome.

## Acknowledgements

We gratefully acknowledge the donation of Metvix cream by Pharmaserve – Lilly S.A.C.I. (Kifisia, Greece) for the needs of this study. The author's G.K. research has been co-financed by Greece and the European Union (European Social Fund- ESF) through the Operational Program «Human Resources Development, Education and Lifelong Learning» in the context of the project "Strengthening Human Resources Research Potential via Doctorate Research" (MIS-5000432), implemented by the State Scholarships Foundation (IKY) [Scholarship number: 2018-050-0502-14578]. The authors E.D. and M.M. gratefully acknowledge the funding of the European Regional Development Fund (ERDF) and National Funds "Synergy ELI-LASERLAB EUROPE, HiPER & IPERION-CH.gr (MIS 5002735)", 2017. The donating and funding sources had no involvement in the study design; in the collection, analysis and interpretation of data; in the writing of the report; and in the decision to submit the article for publication.